\documentclass[aps,preprint,showpacs,epsfig,epsf,superscriptaddress]{revtex4}
\usepackage{graphicx}
\usepackage{amssymb,amsbsy,times}
\usepackage{amsmath}

\begin{document}

\title{Mass-Transport Models with Multiple-Chipping Processes}

\author{Gaurav P. Shrivastav}
\email{gps.jnu@gmail.com}
\affiliation{School of Physical Sciences, Jawaharlal Nehru University,
New Delhi -- 110067, India.} 

\author{Varsha Banerjee}
\email{varsha@physics.iitd.ac.in}
\affiliation{Department of Physics, Indian Institute of Technology,
Hauz Khas, New Delhi -- 110016, India.} 

\author{Sanjay Puri}
\email{puri@mail.jnu.ac.in}
\affiliation{School of Physical Sciences, Jawaharlal Nehru University,
New Delhi -- 110067, India.} 

\begin{abstract}
We study mass-transport models with multiple-chipping processes. The rates of these processes are
dependent on the chip size and mass of the fragmenting site. In this context, we consider $k$-chip
moves (where $k$ $=$ $1, 2, 3,....$); and combinations of $1$-chip, $2$-chip and $3$-chip moves. 
The corresponding mean-field (MF) equations are solved to obtain the steady-state probability 
distributions, $P(m)$ vs. $m$. We also undertake Monte Carlo (MC) simulations of these models. The 
MC results are in excellent agreement with the corresponding MF results, demonstrating that MF 
theory is exact for these models. 

\end{abstract}


\maketitle

\section{Introduction}
\label{sec1}

The formation of structures at the nanoscale has attracted a lot of attention \cite{lagally,z&l}. The 
morphological and statistical properties of these systems are governed by microscopic processes like 
adsorption and desorption, fragmentation, diffusion and aggregation, etc. Different non-equilibrium 
steady-states can be accessed and phase transitions induced if the rates of the these processes are 
varied. Several experimental and theoretical studies have focused on adatom and cluster diffusion
\cite{venables,bartelt,bales,kuipers,takayasu1,takayasu2,sk1}. However, there are relatively few 
studies which consider the fragmentation of clusters, expected to commonly occur during hyperthermal 
ion beam depositions and sputter depositions \cite{rusanen,blandin1,blandin2,roder}. The bombardment of 
clusters by energetic ions induces {\it large island boundary fluctuations} which cause multiple-chipping 
events \cite{rusanen,wucher}. These systems have received limited attention to date because of the 
unavailability of experimental probes to observe fragmentation. Further, the occurrence of these 
events is often on a pico-second time scale, making collection of data difficult. 

The processes of fragmentation and aggregation significantly affect the growth mechanism by 
altering the number of small and large clusters. The steady-state cluster-size distribution in these 
systems is generally characterized by an exponentially decaying tail \cite{rusanen,gpsup}. In contrast, 
systems with adatom and cluster diffusion exhibit steady-state distributions which are power 
laws \cite{takayasu1,sk1,kuiri} . 

In many physical systems, the fragmentation kernel depends upon the masses of the chip size and the 
fragmenting cluster. For example, recent experiments on $Au$ clusters sputtered from embedded $Au$ 
nanoparticles report distinct chipping kernels  for small and large clusters \cite{kuiri}. 
In a related context, groups, herds, schools and flocks of animals also exhibit size-dependent 
fragmentation \cite{bonabeau1,bonabeau2}. Similar observations have been made in the context of 
polymerization \cite{ziff}, gelation \cite{ziff} and complex networks \cite{kwon}. In all these 
systems, the steady-state distributions are either exponentials or power laws or their combinations. 

Fragmentation is thus a ubiquitous phenomenon, observed in a variety of physical systems. In the present 
paper, we study conserved mass models with mass-dependent chipping -- the fragmentation rates depend on 
the chip size and mass of the fragmenting site. We obtain the steady-state mass distributions of these 
models in the mean-field (MF) limit, and compare them with results from Monte Carlo (MC) simulations. 
The purpose of our study is to clarify the generic features of steady-state distributions 
in the presence of multiple-chipping processes.

This paper is organized as follows. In Section 2, we study models with mass-dependent fragmentation and
aggregation processes. We focus on models with $k$-chip moves (where $k$ $=$ $1, 2, 3,...$); and 
combinations of $1$-chip, $2$-chip and $3$-chip moves. In Section 3, we present MC results for 
these models, and compare them with the corresponding MF solutions. We conclude this paper
with a summary and discussion in Section 4.

\section{Transport Models with Mass-Dependent Chipping}
\label{s2} 
Consider a mass-transport model on a discrete lattice where there is no adsorption 
or desorption. To begin with, masses $m_{i}(0)=0,1,2,$ etc. are placed randomly at each site $i$ 
with an overall mass density $\rho$. The evolution of the system is defined by the fragmentation 
kernel $g_{m}(n)$, i.e., the rate for a mass $n$ ($\leqslant m$) to chip from a site with mass $m$.
The $n$-chip then deposits on a randomly-chosen nearest neighbor. The mass of the neighbor adds up, 
while that of the departure site decreases, with the total mass of the system remaining conserved.
Typically, the steady-state  mass distribution [$P(m)$] of sites with mass $m$ in these models 
shows either exponential or power-law decay with $m$.

We study the above model within a MF approximation which keeps track of the distribution of masses, 
ignoring correlations in the occupancy of adjacent sites. Although the MF theory suffers from this 
deficiency, our MC simulations in Sec.~3 show that it gives an accurate description of the above model, 
even in the 1-dimensional case. Let $P(m,t)$ denote the probability that a site has mass $m$ at time $t$. 
In the MF limit, $P(m,t)$ evolves as follows:
\begin{eqnarray}
\label{rate1}
\frac{d}{dt}P(m,t)&=&-P(m,t)\sum_{m_{1}=1}^m g_{m}(m_{1})
      -P(m,t)\sum_{m_{2}=1}^{\infty}P(m_{2},t)\sum_{m_{1}=1}^{m_{2}}
	g_{m_{2}}(m_{1})\nonumber \\
  & &+\sum_{m_{1}=1}^{\infty}P(m+m_{1},t)g_{m+m_{1}}(m_{1})\nonumber\\
      & & +\sum_{m_{1}=1}^{m}P(m-m_{1},t)\sum_{m_{2}=m_{1}}^{\infty} 
    P(m_{2},t)g_{m_{2}}(m_{1}), \ \ \ m \ge 1,
\end{eqnarray}
\begin{eqnarray}
\label{rate10}
\frac{d}{dt}P(0,t)&=&-P(0,t)\sum_{m_{2}=1}^{\infty}P(m_{2},t){\sum_{m_{1}=1}^{m_{2}}}
			g_{m_{2}}(m_{1})          
                  +\sum_{m_{1}=1}^{\infty}P(m_{1},t)g_{m_{1}}(m_{1}).
\end{eqnarray}
Equations (\ref{rate1})-(\ref{rate10}) enumerate all possible ways in which a site with mass $m$ may 
change its mass. The first term on the right-hand-side (RHS) of Eq.~(\ref{rate1}) is the ``loss'' due 
to chipping, while the second term represents the loss due to transfer of mass from a neighbor chipping. 
The third and fourth terms are the ``gain'' terms which represent the ways in which a site with mass 
greater (lesser) than $m$ can lose (gain) the excess (deficit) to yield mass $m$. The terms of 
Eq.~(\ref{rate10}) can be interpreted similarly. The above equations satisfy the sum rule 
\begin{equation}
\label{sumrule}
\frac{d}{dt}\sum_{m=0}^{\infty}P(m,t)=0, \quad \mbox{or} \quad \sum_{m=0}^{\infty}P(m,t) = 1, 
\end{equation}
as required.

Next we consider a specific form of the mass-dependent chipping kernel, $g_m(n)=D(m) n^{-\alpha}$, 
where $\alpha \geqslant 0$ is a parameter. This kernel allows for chips of all sizes, with small masses 
being more likely to fragment from a site than large masses. It is especially relevant in the context 
of fragmentation of sputtered clusters. As revealed by Auger and thermal-desorption spectroscopy 
measurements performed on these nanostructures, fragments of a few atoms display a large mobility
on the surface which rapidly decreases with increasing cluster size \cite{blandin1,davies}.
 
Multiple-chip models may be interpreted as limiting cases of the above model. For $\alpha=\infty,$ this 
kernel only gives rise to $1$-chip processes. For large values of $\alpha$, we expect the fragmentation 
to be dominated by $1$-chip and $2$-chip processes with different rates. As the value of $\alpha$ is 
reduced, the possibility of higher-chip processes becomes appreciable. Finally, in the limit $\alpha=0,$ 
chips of any size are equally likely. Subsequently, we study $1$-chip models, ($1+2$)-chip models, etc. 
as approximations to the model with $g_m(n)=D(m) n^{-\alpha}$. It should be noted that the  
$n^{-\alpha}$-chipping kernel, unlike the multiple-chip models, includes the diffusion move, i.e., the
movement of the entire mass $m$ at a site. The steady-state distribution for this model will be discussed 
in Section III.

\subsection{$k$-chip models and the case $g_m(n)=g(n)$}
\label{s2.1}
First, we study $k$-chip models which act as building blocks to understand models in which chips of 
different sizes are allowed. The chipping kernel has the form
\begin{equation}
\label{kernelk}
g_{m}(n)=w\delta_{n,k}.
\end{equation}
The corresponding rate equations for $P(m,t)$, obtained by substituting Eq.~(\ref{kernelk}) in 
Eqs.~(\ref{rate1})-(\ref{rate10}), are as follows (absorbing $w$ into the time $t$):
\begin{eqnarray}
\label{ratek}
\frac{d}{dt}P(m,t)&=&-(1+s_{k})P(m,t)+P(m+k,t)+s_{k}P(m-k,t), \: m\geq k,\\
\label{ratek0}
\frac{d}{dt}P(m,t)&=&-s_{k}P(m,t)+P(m+k,t), \quad m < k .
\end{eqnarray}
Here, $s_{k}(t)$ = $\sum_{m=k}^{\infty}P(m,t)$ is the probability of sites having mass $k$ or more.

In order to obtain the steady-state solution $P(m)$, the generating function 
$Q(z,t)$ $=$ $\sum_{m=1}^{\infty}z^{m}P(m,t)$ is computed from the above equations, and 
$\partial Q/\partial t$ is set to 0 in the steady state. $P(m)$ is then obtained by evaluating the integral
\begin{eqnarray}
\label{steady1}
P(m)&=&\frac{1}{2\pi i}\int_{C}dz~\frac{Q(z)}{z^{m+1}},\hspace{0.3cm} m>1,
\end{eqnarray}
where the contour $C$ encircles the origin in the complex plane. The steady-state generating function of the 
$k$-chip model is
\begin{equation}
\label{genk}
Q(z) = \frac{z(s_{1}-s_{2})+z^{2}(s_{2}-s_{3})+\cdot\cdot\cdot\cdot+z^{k}s_{k}(1-s_{1})}
        {(1-s_{k}z^{k})}.
\end{equation}

The conservation of mass requires that $\sum_{m=1}^{\infty} mP(m)$ = $\rho$, where $\rho$ is the mass density.
Putting $dQ/dz\big|_{z=1}=\rho$, we obtain 
\begin{equation}
\label{rhok}
\rho=\frac{s_{1}+s_{2}+.......+s_{k}}{1-s_{k}}.
\end{equation}
The corresponding $P(m)$, evaluated from Eq.~(\ref{steady1}), comprises of $k$ branches \cite{sbp10}: 
\begin{eqnarray}
\label{steadym}
P(m)&=& \sum_{i=0}^{k-1}
		(s_{i}-s_{i+1}) s_k^{(m-i)/k}\delta_{\mbox{mod}(m,k),i}.
\end{eqnarray}
(We have defined $s_{0}$ = 1 in the above equation.) Thus all the branches decay exponentially.
The occupation probability of the $i^{\mbox{th}}$ branch ($i$ $=$ $0\to k-1$) is $ S_{i}=P(i)+P(i+k)+P(i+2k)+...$. 
Notice that because of the nature of the $k$-chip moves, the initial population $S_{i}$ in each of the 
branches is conserved at all times. Therefore, there are $k$ conserved quantities in 
addition to the conserved mass. These enable us to determine the $s_{i}$'s in terms of 
$\rho$ and the $S_{i}$'s.

We briefly discuss the solution of the 1-chip model because of its 
ubiquitous nature. On substituting $k$ $=$ 1 in Eq.~(\ref{steadym}), the steady-state distribution
 simplifies to \cite{sk3,rajesh}.
\begin{eqnarray}
\label{soln1}
P(m) &=& s_{1}^{m} - s_{1}^{m+1}=\frac{1}{1+\rho}\left( \frac{\rho}{1+\rho}\right)^{m}
\equiv a_{1}b_{1}^{m}.
\end{eqnarray}
This 1-chip solution is actually valid for a wide range of mass transport models. It
arises whenever the fragmentation kernel $g_m(n)=g(n)$, i.e., the chipping rate is independent 
of the mass of the departure site \cite{sbp10}. This can be verified by substituting Eq.~(\ref{soln1})
in Eqs.~(\ref{rate1})-(\ref{rate10}). Further, the above kernels can be written as a product of two 
non-negative functions, i.e., $g_m(n)= f(n)h(m-n)/h(m)$ where $h(x)$ is a constant here. They therefore 
satisfy the necessary and sufficient condition for the steady-state distributions to become factorizable. 
Evans et al. have shown that mean field theory is exact for this class of models \cite{emz06}. 

\subsection{(1+2)-chip model}
\label{s2.2}
We next consider a combination of 1-chip and 2-chip processes. Our MC simulations show that these
mimic the $g_m(n)=D(m) n^{-\alpha}$ model for $\alpha \gtrsim 4$. We consider the general
(1+2)-chip model with the chipping kernel:
\begin{eqnarray}
\label{12gen}
 g_{m}(n)&=&w_{1}\delta_{n,1}\delta_{m,1}+(w_{2}\delta_{n,1} + w_{3}\delta_{n,2})\theta(m-2) , 
\end{eqnarray}
where $w_{1}, w_{2}, w_{3}$ are the respective chipping rates and $\theta(x) = 1$ for $x\geqslant 0$. Notice that 
the above kernel has an explicit $m$-dependence, i.e., the 1-chip solution is not a steady-state solution for the 
corresponding rate equations except in special cases.   

Replacing $g_{m}(n)$ in Eqs.~(\ref{rate1})-(\ref{rate10}),  we obtain the following rate equations:
\begin{eqnarray}
\label{rate12g}
\frac{d}{dt}P(m,t)&=&-[w_1( s_1-s_2)+(w_2+w_3)(1+s_2)]P(m,t)+w_2 P(m+1,t) \nonumber\\
& & +w_3 P(m+2,t)+ \left[w_1(s_1-s_2)+w_2 s_2\right] P(m-1,t)\nonumber\\
& &+w_3s_2P(m-2,t),\quad m\geq 2, \\
\label{rate12g1}
\frac{d}{dt}P(1,t)&=&-[w_1(1+s_1-s_2)+(w_2+w_3)s_2] P(1,t)+ w_2 P(2,t)+w_3 P(3,t) \nonumber \\
& & + [w_1(s_1-s_2)+w_2 s_2]P(0,t), \\
\label{rate12g0}
\frac{d}{dt}P(0,t)&=&-[w_1(s_1-s_2)+(w_2+w_3)s_2]P(0,t) + w_1 P(1,t) + w_3 P(2,t).
\end{eqnarray}

As usual, we are interested in the steady-state solution of this model. The steady-state generating
 function can be obtained using Eqs.~(\ref{rate12g})-(\ref{rate12g0}), and some algebra yields 
\begin{eqnarray}
\label{qz12}
Q(z)&=&\frac{N(z)}{D(z)},\\
N(z)&=&w_{3}s_{2}(1-s_{1})z^{3}+[w_{2}s_{1}(1-s_{2})-w_{1}s_{1}(s_{1}-s_{2})+w_{3}s_{1}(1-s_{2})]z^{2}\nonumber\\
&&+w_{3}(s_{1}-s_{2})z,\\
D(z)&=&-w_{3}s_{2}z^{3}-\left[w_{2}s_{2}+w_{1}(s_{1}-s_{2})+w_{3}s_{2}\right]z^{2}+(w_{2}+w_{3})z+w_{3}.
\end{eqnarray}
Further, the relation between the mass density $\rho$ and $s_{1}$, $s_{2}$ is obtained from
$dQ(z)/dz \big|_{z=1}=\rho$ as follows:
\begin{equation}
\label{rho12}
\rho=\frac{w_{2}s_{1}+2w_{3}(s_{1}+s_{2})}{w_{2}(1-s_{2})+2w_{3}(1-s_{2})-w_{1}(s_{1}-s_{2})}.
\end{equation}

To obtain $P(m)$, we need to invert $Q(z)$ using Eq.~(\ref{steady1}). This integral is done by finding
the roots of the numerator $N(z)$ (which is of the form $z \times$ quadratic in $z$) and
the denominator $D(z)$ (which is cubic in $z$), and rewriting $Q(z)$ in the form of partial fractions. 
This procedure often proves to be very cumbersome. An easier route is to solve the recurrence 
relation for $P(m)$, which can be obtained directly from Eqs.~(\ref{rate12g})-(\ref{rate12g0}) 
by setting the LHS to zero in the steady state. It may also be obtained by comparing the 
coefficients of the $z^{m}$- terms on either side of Eq.~(\ref{qz12}). Some algebra yields the 
recurrence relation:
\begin{eqnarray}
\label{genrec12}
P(m)&=&-\frac{(w_{2}+w_{3})}{w_3}P(m-1)+\frac{s_{2}(w_{2}+w_{3})+w_{1}(s_{1}-s_{2})}{w_3}P(m-2)\nonumber\\
&&+ s_{2}P(m-3),  \quad m\geq 3.
\end{eqnarray}
To obtain the solution, we assume that 
\begin{equation}
\label{expo}
P(m) = Ax^{m}. 
\end{equation}
Substituting this form in Eq.~(\ref{genrec12}) results in a cubic equation for $x$:
\begin{eqnarray}
\label{cube}
x^{3}+\left( 1+\frac{w_{2}}{w_{3}}\right)x^2-\left[s_{2}
	+\frac{w_{2}s_{2}+w_{1}(s_{1}-s_{2})}{w_{3}}\right]x-s_{2}=0.
\end{eqnarray}
The roots of this cubic equation are denoted as $x_{1}, x_{2}, x_{3}$, and are all real in this case \cite{a&s}. 
(For brevity, we do not present their explicit forms.)

The steady-state solution may then be written as 
\begin{eqnarray}
\label{gensol}
P(m)=A_{1}x_{1}^{m}+A_{2}x_{2}^{m}+A_{3}x_{3}^{m}.
\end{eqnarray}
Notice that the recurrence relation in Eq.~(\ref{genrec12}) is valid for $m\ge 3$. The following choice of
the coefficients $A_{1}$, $A_{2}$ and $A_{3}$ ensures that $P(0)$, $P(1)$ and $P(2)$ also follow Eq.~(\ref{gensol}):
\begin{eqnarray}
A_{1}&=&\frac{(1-s_{1})x_{2}x_{3}-(s_{1}-s_{2})(x_{2}+x_{3})+(s_{2}-s_{3})}{(x_{1}-x_{2})(x_{1}-x_{3})},\\
A_{2}&=&\frac{-(1-s_{1})x_{1}x_{3}+(s_{1}-s_{2})(x_{1}+x_{3})-(s_{2}-s_{3})}{(x_{1}-x_{2})(x_{2}-x_{3})},\\
A_{3}&=&\frac{(1-s_{1})x_{1}x_{2}-(s_{1}-s_{2})(x_{1}+x_{2})+(s_{2}-s_{3})}{(x_{1}-x_{3})(x_{2}-x_{3})}.
\end{eqnarray}

Of the three roots, we find that $\mid x_1 \mid, \mid x_2 \mid < 1$ and $\mid x_3 \mid > 1$. In order to ensure 
that $P(m)$ in Eq.~(\ref{gensol}) is meaningful, we set $A_3=0$. This condition, together with Eq.~(\ref{rho12}), 
can be solved self-consistently to obtain $s_1$ and $s_{2}$ in terms of $w_{1},w_{2},w_{3}$ and $\rho$. The 
coefficients $A_{1}$, $A_{2}$ are thereby determined, and the steady-state distribution of the ($1+2$)-chip 
model is thus a combination of two power-law (exponential) functions:
\begin{equation}
\label{steady12}
P(m)=A_{1}x_{1}^{m}+A_{2}x_{2}^{m}.
\end{equation}
As usual, the large-$m$ behavior is dominated by the slower of the two exponentials. 

An alternative approach is to obtain $A_{1}$ and $A_{2}$ in terms of ($x_{1}$, $x_{2}$, $x_{3}$, $s_{1}$, $s_{2}$) 
from the equations:
\begin{eqnarray}
\label{p0}
P(0) &=& 1-s_{1} = A_{1}+A_{2},\\
\label{p1}
P(1) &=& s_{1}-s_{2} = A_{1}x_{1}+A_{2}x_{2}.
\end{eqnarray}
We can then treat ($x_{1}$, $x_{2}$, $x_{3}$, $s_{1}$, $s_{2}$) as unknowns to be determined by 
$5$ coupled equations as follows:\\
(a) The $x_{i}$'s must satisfy the cubic equation (\ref{cube}).\\
(b) The normalization condition provides the constraint
\begin{equation}
\label{norm}
1= \frac{A_{1}}{1-x_{1}}+\frac{A_{2}}{1-x_{2}}.
\end{equation}
(c) Equation~(\ref{rho12}) for $\rho$.\\
These equations can be solved numerically to obtain the unknowns in terms of $w_{1}$, $w_{2}$, $w_{3}$, $\rho$ 
and thereby the solution.

It is useful to consider limits where the double-exponential function in Eq. (\ref{steady12}) reverts to a 
single-exponential form. \\
\noindent (i) $w_3=0$\\
If we substitute $w_{3}=0$ in Eq.~(\ref{12gen}), the chipping kernel simplifies to
\begin{equation}
\label{case1}
g_{m}(n)=w_1\delta_{n,1}\delta_{m,1}+w_2\delta_{n,1}\theta(m-2).
\end{equation}
This model has only 1-chip processes, but the 1-chip rate is different for sites with one 
unit of mass ($m=1$) and those with two or more units of mass ($m \geq 2$). In this limit, the recurrence 
relation in Eq.~(\ref{genrec12}) reduces to
\begin{eqnarray}
\label{rec12R}
P(m)&=&\left[s_{2}+\frac{w_{1}}{w_{2}}(s_{1}-s_{2})\right]P(m-1), \quad m\ge 2.
\end{eqnarray}

The resultant steady-state distribution has the following form:
\begin{eqnarray}
\label{sscase1}
P(m) =\left\{\begin{array}{ll}
a_{2}b_{2}^{m}, &\mbox {$m \geqslant 1$},\\
1-s_1,            & \mbox{$m=0$}, \\
\end{array}
\right.
\end{eqnarray}
where
\begin{equation}
\label{a2}
a_{2}=\frac{w_2(s_1 - s_2)}{w_1(s_1-s_2)+w_2 s_2},
\quad b_{2}=\frac{w_1(s_1-s_2)+w_2 s_2 }{w_2}.
\end{equation}
We obtain $s_{1}$, $s_{2}$ as functions of $w_{1}$, $w_{2}$, $\rho$:
\begin{eqnarray}
\label{s1s2}
s_1 &=& \frac{-w_{2}(1+\rho)+\sqrt{w_{2}^{2}(1-\rho)^{2}+4w_{1}w_{2}\rho}}{2(w_{1}-w_{2})},\\
s_2 &=& \frac{w_{1}s_{1}^{2}}{w_{2} + (w_{1}-w_{2})s_{1}}.
\end{eqnarray}

The exponentially-decaying solution in Eq.~(\ref{sscase1}) has a different slope from that in Eq. (\ref{soln1}). 
In the limit $w_1=w_2$, the solution in Eq.~(\ref{sscase1}) reduces to the 1-chip solution of 
Eq.~(\ref{soln1}). \\
\noindent (ii) $w_1=w_{2}$, $w_{3}$ arbitrary\\
If we substitute $w_1=w_2$ in Eq.~(\ref{12gen}), the chipping kernel reduces to  
\begin{eqnarray}
\label{case11}
g_{m}(n) &=& (w_{1}\delta_{n,1} +w_{3}\delta_{n,2})\theta(m-n).
\end{eqnarray}
This kernel has the functional form $g_{m}(n)=g(n)$, and
the corresponding $P(m)$ is the $1$-chip solution in Eq.~(\ref{soln1}).

\subsection{(1+2+3)-chip model}
\label{s2.3}
We now consider a model with the possibility of chipping 1, 2 or 3 units of mass: 
\begin{eqnarray}
\label{123gen}
g_{m}(n)&=&w_{1}\delta_{n,1}\delta_{m,1}
+(w_{2}\delta_{n,1}+w_{3}\delta_{n,2})\delta_{m,2}+(w_{4}\delta_{n,1}+w_{5}\delta_{n,2}
	+w_{6}\delta_{n,3})\theta(m-3).
\end{eqnarray}
Substituting Eq.~(\ref{123gen}) in Eqs.~(\ref{rate1})-(\ref{rate10}) yields the required rate equations, 
which we do not present here. Some algebra yields the generating function: 
\begin{eqnarray}
\label{q123gen}
Q(z) &=&\frac{\overline{N}(z)}{\overline{D}(z)},\\
\overline{N}(z)&=&w_{6}s_{3}(1-s_{1})z^{5}+\big[-(w_{2}+w_{3}s_{1})(s_{2}-s_{3})+w_{4}(s_{2}-s_{3})\nonumber\\
&&+(w_{5}+w_{6})(s_{2}-s_{1}s_{3})\big]z^{4}+\big[-w_{1}s_{1}(s_{1}-s_{2})-(w_{2}+w_{3})s_{1}(s_{2}-s_{3})
\nonumber\\
&&+(w_{2}-w_{4})(s_{2}-s_{3})+s_{1}(w_{4}+w_{5}+w_{6})\big]z^{3}+\left[w_{5}(s_{1}-s_{2})+w_{6}(s_{2}-s_{3})\right]z^{2}\nonumber\\
&&+w_{6}(s_{1}-s_{2})z,\\
\overline{D}(z)&=&-w_{6}s_{3}z^{5}-[w_{3}(s_{2}-s_{3})+(w_{5}+w_{6})s_{3}]z^{4}-[w_{1}(s_{1}-s_{2})+(w_{2}+w_{3})(s_{2}-s_{3})\nonumber\\
&&+(w_{4}+w_{5}+w_{6})s_{3}]z^{3}+(w_{4}+w_{5}+w_{6})z^{2}+(w_{5}+w_{6})z+w_{6}.
\end{eqnarray}

As before, the steady-state distribution is obtained from the recurrence relation for $P(m)$:
\begin{eqnarray}
\label{genrec123}
w_{6}P(m)&=&-(w_{5}+w_{6})P(m-1)-(w_{4}+w_{5}+w_{6})P(m-2)\nonumber\\
&&+\left[w_{1}(s_{1}-s_{2})+(w_{2}+w_{3})(s_{2}-s_{3})+(w_{4}+w_{5}+w_{6})s_{3}\right]P(m-3)\nonumber\\
&&+\left[w_{3}(s_{2}-s_{3})+(w_{5}+w_{6})s_{3}\right]P(m-4)+s_{3}w_{6}P(m-5),\hspace{0.1cm}m\geqslant 5. 
\end {eqnarray}
Substituting $P(m)=Bx^{m}$ in the above equation results in the following quintic equation:
\begin{eqnarray}
\label{quintic}
&&x^{5}+\left(1+\frac{w_{5}}{w_{6}}\right)x^{4}+\left(1+\frac{w_{4}+w_{5}}{w_{6}}\right)x^{3}\nonumber\\
&&-\left[\frac{w_{1}}{w_{6}}(s_{1}-s_{2})+\frac{w_{2}+w_{3}}{w_{6}}(s_{2}-s_{3})+\left(1+\frac{w_{4}+w_{5}}{w_{6}}\right)s_{3}\right]x^{2}\nonumber\\
&&-\left[\left(1+\frac{w_{5}}{w_{6}}\right)s_{3}+\frac{w_{3}}{w_{6}}(s_{2}-s_{3})\right]x-s_{3}=0.
\end{eqnarray}
This quintic equation cannot be solved explicitly in terms of radicals. However, we can generalize the alternative 
approach described for the ($1+2$)-chip model subsequent to Eq.~(\ref{steady12}). The resultant coupled equations 
(solved numerically) yield the required solution for given values of $w_{1}$ to $w_{6}$ and $\rho$. Typically, 
$P(m)$ for the (1+2+3)-chip model is a sum of three exponential functions. 

As before, it is instructive to examine some simple limits of this model.\\
\noindent (i) $w_{3},w_{5}$, $w_{6}=0$\\
This model has only 1-chip processes, but the chipping rates depend on the mass of the departure site:
\begin{eqnarray}
\label{123s2}
g_{m}(n)&=&w_{1}\delta_{n,1}\delta_{m,1}+w_{2}\delta_{n,1}\delta_{m,2}+w_{4}\delta_{n,1}\theta(m-3).
\end{eqnarray}
The corresponding distribution can be obtained from the simplified version of
 Eq.~(\ref{genrec123}), or the simplified $Q(z)$ from Eq.~(\ref{q123gen}):
\begin{eqnarray}
\label{123s2ss}
P(m) = \left\{\begin{array}{ll}
a_{3} b_{3}^{m}, &\mbox {$m \ge 2$},\\
s_{1}-s_{2}, & \mbox{$m=1$},\\
1-s_1,            & \mbox{$m=0$}, \\
\end{array}
\right.
\end{eqnarray}
where
\begin{eqnarray}
\label{a3}
a_{3}=\frac{w_{4}^{2}(s_2 - s_3)}{[w_1(s_1-s_2)+ w_2(s_2-s_3)]^{2}+w_4 s_3 },\quad
 b_{3}=\frac{ w_1(s_1-s_2) +  w_2(s_2-s_3) + w_4 s_3 }{w_4}.
\end{eqnarray}

 These results, along with the equation for $\rho$, enable us to determine $s_{i}$'s 
in terms of $w_{i}$'s and $\rho$. It is straightforward to generalize Eqs.~(\ref{sscase1})-(\ref{a2}) and 
Eqs.~(\ref{123s2ss})-(\ref{a3}) to the $1$-chip model with different rates for departure-site masses $m=1$, 
$m=2$,......., $m\geqslant k$.\\
\noindent (ii) $w_{6}=0$\\
With this substitution, the chipping kernel of Eq.~(\ref{123gen}) reduces to a generalized version of the 
($1+2$)-chip model discussed earlier: the $1$-chip and $2$-chip rates 
are now different for sites with $m=1$, $m=2$ and $m\geqslant 3$. The recurrence relation in this case also yields 
a cubic equation. The steady-state probability distribution has the same form as Eq.~(\ref{steady12}), but the 
decay rates of the two exponential functions are distinct from those in Eq.~(\ref{steady12}). In 
the limit $w_{2}=w_{4}$ and $w_{3}=w_{5}$, we recover Eq.~(\ref{steady12}).\\
\noindent(iii) $w_{1}=w_{2}=w_{4}$ and $w_{3}=w_{5}$\\
This case corresponds to the chipping rates being independent of the site mass. In this limit, the 
chipping kernel in Eq.~(\ref{123gen}) becomes
\begin{eqnarray}
\label{123s3}
g_{m}(n)&=&(w_{1}\delta_{n,1}+w_{3}\delta_{n,2}+w_{6}\delta_{n,3})\theta(m-n).
\end{eqnarray}
Hence, the steady-state distribution in this limit is the $1$-chip solution of Eq.~(\ref{soln1}). 
\section{Monte Carlo Simulations}
\label{s3}

In this section, we present Monte Carlo (MC) results for some of the models discussed earlier. All simulations were 
performed on 1-$d$ or 2-$d$ lattices with periodic boundary conditions. The lattice sizes were $L=1024$ in $d=1$, 
and $L^{2}=128^{2}$ in $d=2$. The initial condition for a run consists of a random distribution of masses 
with density $\rho$. We evolve the system using the chipping rate $g_{m}(n)$, and compute the mass distribution 
$P(m)$. This quantity settles to equilibrium for $t>25000$ MCS - we show results for $P(m)$ vs. $m$ at $t=50000$ MCS. 
The statistical data presented here was obtained as an average over $200$ independent runs.

First, we present results for the kernel $g_{m}(n)= g(n)$, discussed in Sec.~2.1. In Fig. 1, we plot 
$P(m)$ vs. $m$ obtained from 1-$d$ MC simulations with $\rho = 5$ and three different functional forms of 
$g_{m}(n)=1, 1/n, n^{3}e^{-0.2n}$ and $\rho=5$. The MC data sets are numerically coincident with each other. They 
are also in excellent agreement with the 1-chip solution in Eq.~(\ref{soln1}) (denoted as a solid line), which was 
obtained from the corresponding MF equations. Our subsequent results also show that the MC data is described very 
well by the solutions of the corresponding MF equations, even for $d=1$. This demonstrates that the MF equations 
are exact in the present context \cite{emz06}.

Second, we present results for the (1+2)-chip model discussed in Sec.~2.2. Fig. 2 shows the steady-state 
distributions obtained from 1-$d$ and 2-$d$ MC simulations with $\rho=5$, and $w_{1}=6, w_{2}=0.5$ and $w_{3}=7$. 
The solid line denotes the result in Eq.~(\ref{steady12}) - the corresponding roots and coefficients are specified 
in the caption. The oscillatory structure of $P(m)$ arises as one of the roots is negative. For the wide range of 
parameter values we considered, one of the roots is always found to be negative.

Next, let us consider the arbitrary kernel $g_{m}(n)=D(m)/n^{\alpha}$ with $D(m)$ determined from 
$\sum_{n=1}^{m}D(m)/n^{\alpha}=1$. Unlike our earlier models, this fragmentation kernel allows  
diffusion, which corresponds to the movement of the entire mass $m$ at a site to a randomly chosen
neighbor. In Fig. 3, we show $P(m)$ vs. $m$ from MC simulations in 
$d=1,2$ with $\alpha=3$ and $\rho=5$. The solid line denotes the analytical result for the analogous 
($1+2+3$)-chip model with $w_{i}$'s specified in the caption. Clearly, the ($1+2+3$)-chip approximation 
captures the original model rather well -- this is true for $\alpha \gtrsim 2$. 

Finally, in Fig. 4, we show $P(m)$ vs. $m$ from MC simulations in $d=1,2$ for the $1$-chip model with different 
chip rates upto $m\geqslant 5$, i.e., $g_{m}(n)=\delta_{n,1}[w_{1}\delta_{m,1}+w_{2}\delta_{m,2}+w_{3}\delta_{m,3}
+w_{4}\delta_{m,4}+w_{5}\theta(m-5)]$. We show results for $\rho=5$ and the $w_{i}$'s are specified 
in the caption. The solid line denotes the analytical solution obtained by generalizing the solution in 
Eqs.~(\ref{123s2ss})-(\ref{a3}). This solution is exponential for $m\geqslant 4$.

\section{Summary and Discussion}
\label{s4}

We conclude this paper with a summary and discussion of the results presented here. We have studied the 
steady-state distributions [$P(m)$ vs. $m$] for mass-transport models with multiple-chipping processes which 
depend upon the masses of the departure sites and the chips. These models are relevant for a variety of 
physical applications. In general, a site with mass $m$ could have 1, 2, 3,.., $m$ units of mass chip with 
different rates and aggregate with a neighbor. In this context, we study $k$-chip processes 
(where $k$ $=$ 1,2,3,...) and combinations thereof. We undertake Monte Carlo (MC) simulations of these models, 
and analytically study the steady-state solutions of the corresponding mean-field (MF) equations. Our MC results 
are in excellent agreement with the MF solutions, demonstrating that MF theory is exact in this context. This
is true even when the kernels are not factorizable \cite{emz06}, and requires further investigation.

The steady-state distribution of the $k$-chip models has $k$ branches, each of which decays exponentially with the 
same slope. We find that a large class of chipping kernels, where $g_m(n)$ is independent of $m$, gives rise to an 
exponentially-decaying distribution: $P(m)=(1+\rho)^{-1}[\rho/(1+\rho)]^{m}$, where $\rho$ is the mass density.
This is also the MF solution for the 1-chip model, where one unit of mass fragments from a site and aggregates 
with a randomly-chosen nearest-neighbor.

We have also studied models with a combination of 1-chip, 2-chip and 3-chip processes. The steady-state 
distributions of the $(1+2)$-chip model and $(1+2+3)$-chip model are sums of two exponential functions and three 
exponential functions, respectively. We expect the steady-state distribution of the $(1+2+\ldots +k)$-chip model 
to be a sum of $k$ exponentially-decaying functions.  We have also examined several limiting cases of the above 
models. Our conclusion is that the steady-state distribution is sensitive to slight changes in the chipping kernel.

Finally, it is relevant to discuss physical processes which give rise to multi-exponential mass
distributions vs. power-law distributions. Power laws have been observed in a class of mass-transport 
models where single-particle adsorption or chipping processes and (mass-independent) diffusion processes 
maintain a delicate balance between the lower and upper ends of the mass spectrum 
\cite{takayasu1,takayasu2,sk1}. On the other hand, mass-dependent fragmentation and aggregation precludes 
this balance. Consequently, all our models exhibit exponential distributions. These would be relevant in the 
context of ion-beam and sputter-deposited nanostructures, animal group distributions, 
polymerization, gelation and complex networks.\\

\noindent{\bf Acknowledgments}

GPS and VB would like to acknowledge the support of CSIR Grant No. 03(1077)/06/EMR-II.

\newpage
\begin{figure}[htb!]
\centering
\includegraphics[width=15.0cm,height=20.0cm,angle=270]{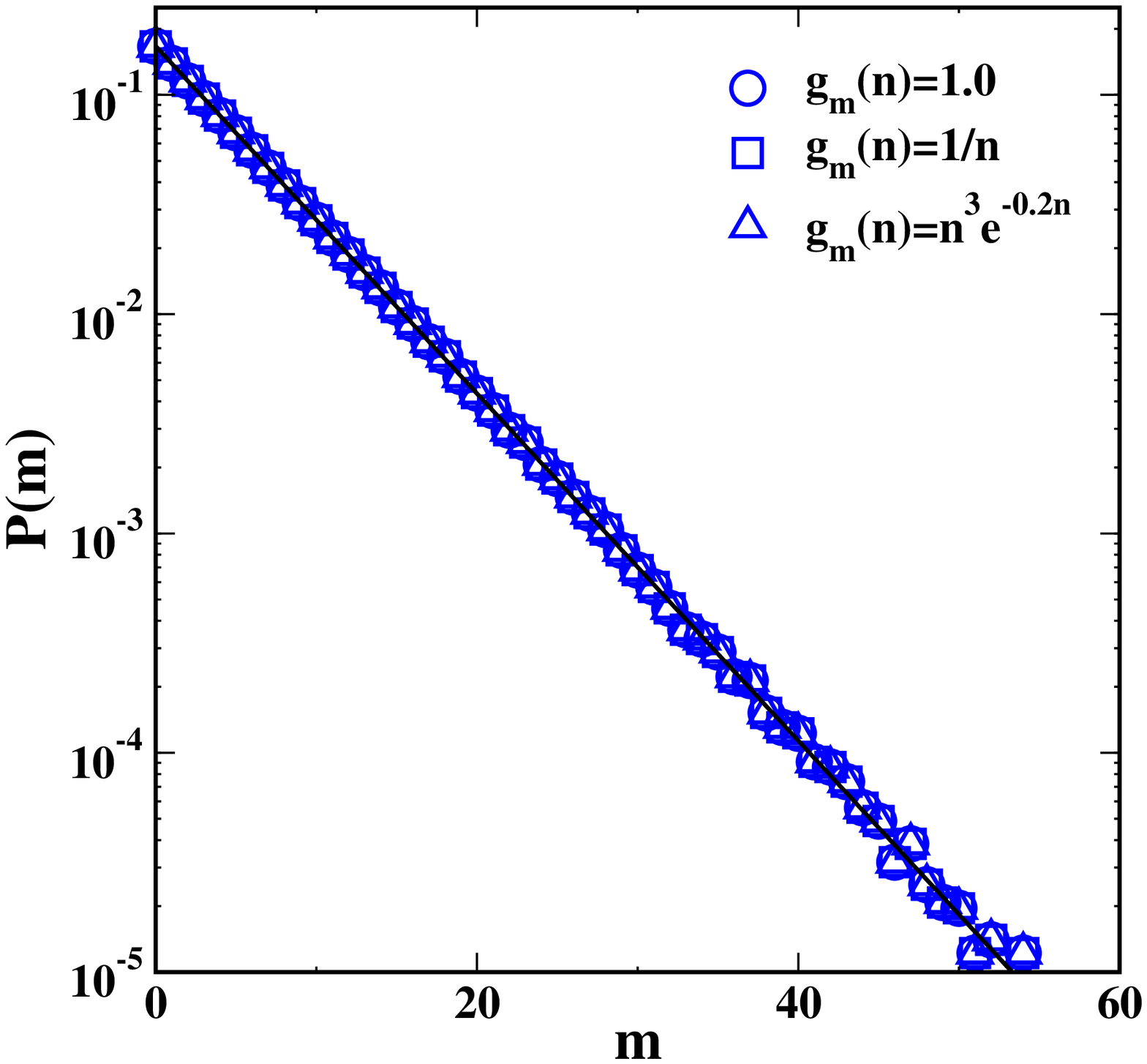}
\caption{Steady-state probability distributions [$P(m)$ vs. $m$] from
$d=1$ MC simulations with three different forms of $g_m(n)$. The data sets are plotted on a linear-logarithmic scale. The details of the MC simulations are provided in the text. The mass density is $\rho = 5$. The solid line denotes the 1-chip solution in Eq.~(\ref{soln1}).}
\label {Figure 1}
\end{figure}
\newpage

\begin{figure}[htb!]
\centering
\includegraphics[width=15.0cm,height=20.0cm,angle=270]{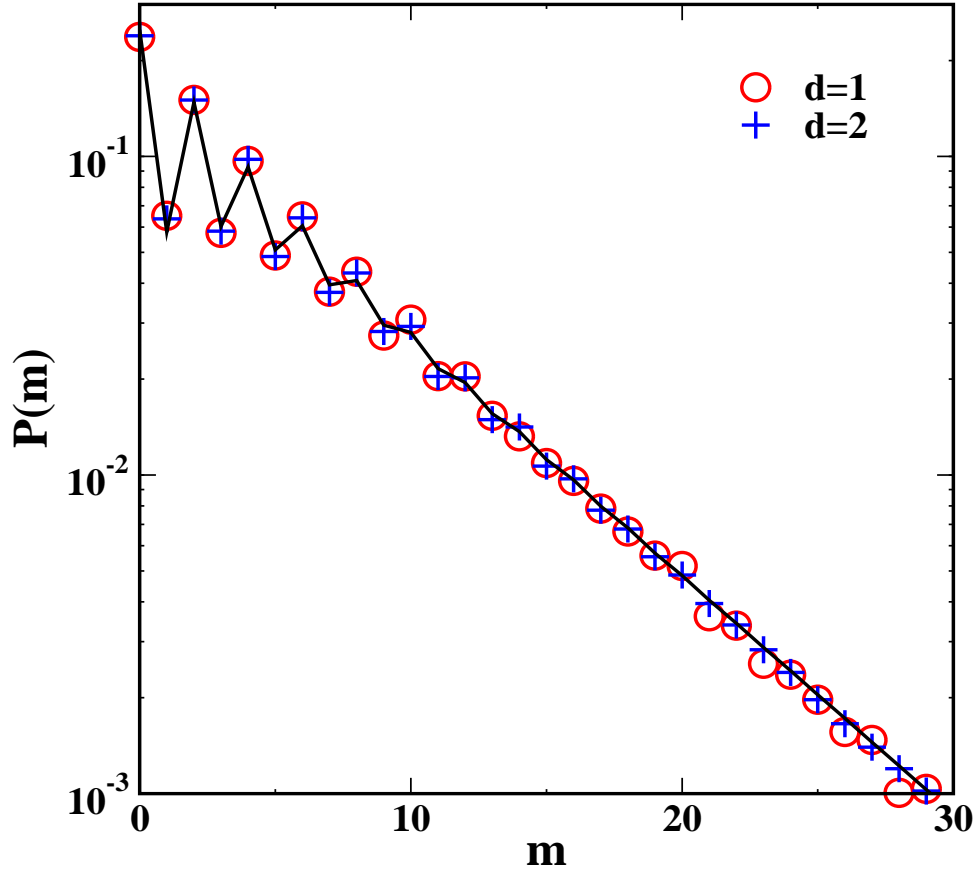}
\caption{Plot of $P(m)$ vs. $m$ for the (1+2)-chip model, obtained from MC simulations in $d=1,2$ with $\rho = 5$. We used $g_m(n)$ in Eq.~(\ref{12gen}) with $w_1=6, w_2 =0.5$ and $w_3=7$. The solid line denotes the solution in Eq.~(\ref{steady12}) with $x_1=0.8427$, $x_2=-0.6498$, $A_1=0.1475$, $A_2=0.1021$.}
\label{Figure 2}
\end{figure}

\newpage

\begin{figure}[htb!]
\centering
\includegraphics[width=15.0cm,height=20.0cm,angle=270]{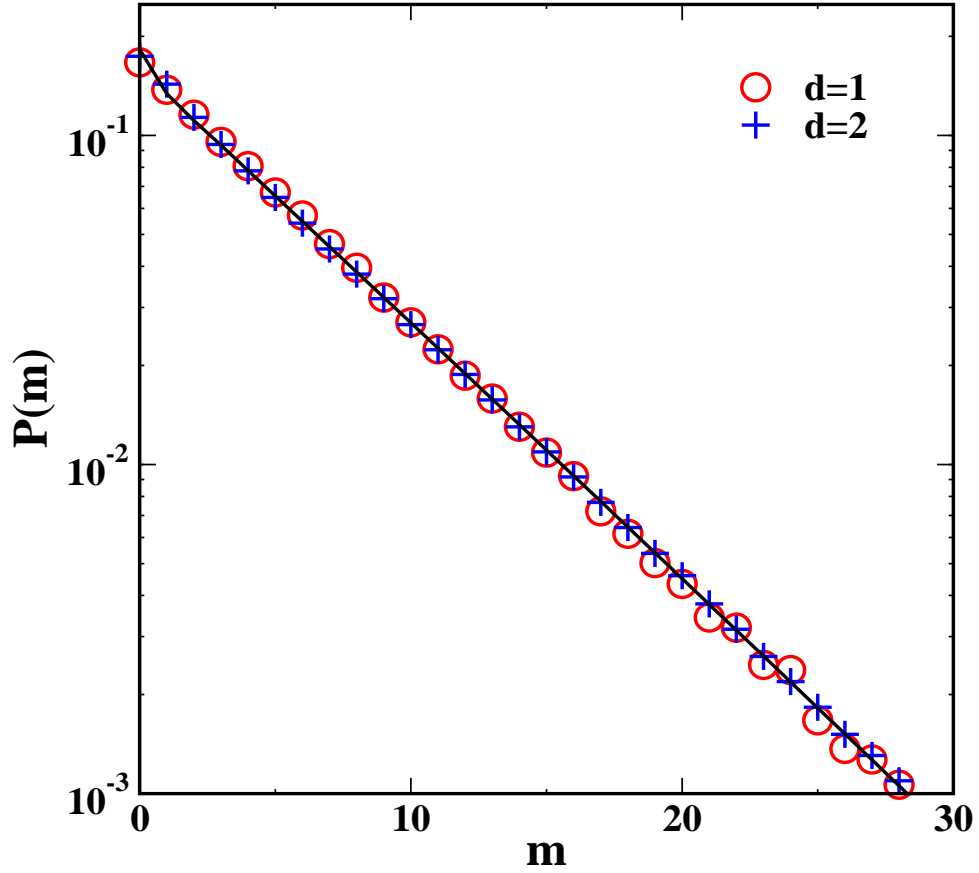}
\caption{Plot of $P(m)$ vs. $m$, obtained from MC simulations in  $d=1,2$ with kernel $g_{m}(n)=D(m)/n^{\alpha}$ [where $D(m)^{-1}=\sum_{n=1}^{m}n^{-\alpha}$ ] for $\alpha=3$ and $\rho=5$. The solid line denotes the result for the analogous (1+2+3)-chip model with  $w_{1}=1$, $w_{2}=8/9$, $w_{3}=1/9$, $w_{4}=216/251$, $w_{5}=27/251$ and $w_{6}=8/251$. The solution is a sum of three exponentials with $x_1=0.8372$, $x_2=-0.0594-0.1263i$, $x_3=-0.0594+0.1263i$, $A_1=0.1590$, $A_2=0.0118+0.0090 i$, $A_3=0.0118-0.0090i$.}
\label{Figure 3}
\end{figure}
\newpage

\begin{figure}[htb!]
\centering
\includegraphics[width=15.0cm,height=20.0cm,angle=270]{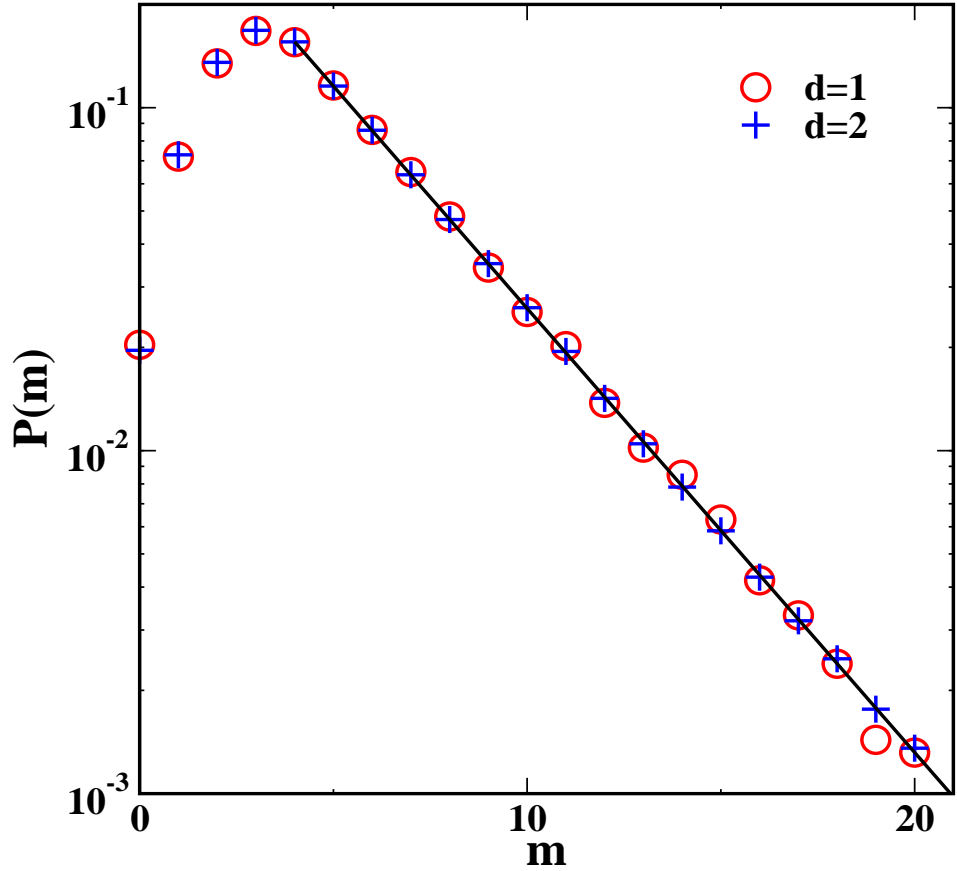}
\caption{Plot of $P(m)$ vs. $m$ for the $1$-chip model with different chip rates upto $m\geqslant 5$.
The symbols denote MC simulation results in $d=1,2$ with $\rho=5$ and $w_{1}=1$, $w_{2}=2$, $w_{3}=3$, $w_{4}=4$, $w_{5}=5$. The solid line denotes the generalization of Eqs.~(\ref{123s2ss})-(\ref{a3}). The solution decays exponentially for $m\geqslant 4$, $P(m)=0.5133\times(0.7421)^{m}$.}
\label{Figure 4}
\end{figure}


\begin{thebibliography}{99}
\bibitem{gpsup}
We have checked this by fitting exponential distributions to the experimental data in \cite{blandin2}.
\end{thebibliography}
\end{document}